# High-Q nanophotonics over the full visible spectrum enabled by hexagonal boron nitride metasurfaces

Lucca Kühner[1], Luca Sortino[1], Benjamin Tilmann[1], Thomas Weber[1], Kenji Watanabe[2], Takashi Taniguchi[3], Stefan A. Maier[4,5,1], and Andreas Tittl[1,*]

[1]Chair in Hybrid Nanosystems, Nanoinstitute Munich, and Center for NanoScience, Faculty of Physics, Ludwig-Maximilians-University Munich, Königinstraße 10, 80539 München, Germany

[2]Research Center for Functional Materials, National Institute for Materials Science, 1-1 Namiki, Tsukuba 305-0044, Japan

[3]International Center for Materials Nanoarchitectonics, National Institute for Materials Science, 1-1 Namiki, Tsukuba 305-0044, Japan

[4]School of Physics and Astronomy, Monash University, Wellington Rd, Clayton VIC 3800, Australia

[5]The Blackett Laboratory, Department of Physics, Imperial College London, London, SW7 2AZ, United Kingdom

E-mail: andreas.tittl@physik.uni-muenchen.de

**Abstract**

All-dielectric optical metasurfaces with high quality ($Q$) factors have so far been hampered by the lack of simultaneously lossless and high refractive index (RI) materials over the full visible spectrum. To achieve broad spectral coverage, the use of low-index materials is, in fact, unavoidable due to the inverse correlation between the band-gap energy (and therefore the optical losses) and the RI. However, for Mie resonant photonics, smaller RIs are associated with reduced $Q$ factors and mode volume confinement. In this work, we leverage symmetry-broken bound states in the continuum (BICs) to efficiently suppress radiation losses from the low-index (n~2) van der Waals material hexagonal boron nitride (hBN), realizing metasurfaces with high-$Q$ resonances over the complete visible spectrum. In particular, we analyze the rational use of low and high RI materials as resonator components and harness our insights to experimentally demonstrate sharp BIC resonances with $Q$ factors above 300, spanning wavelengths between 400 nm and 1000 nm from a single hBN flake. Moreover, we utilize the enhanced electric near-fields to demonstrate second harmonic generation (SHG) with enhancement factors above $10^2$. Our results provide a theoretical and experimental framework for the implementation of low RI materials as photonic media for metaoptics.

**Introduction**

All-dielectric materials have recently emerged in nanophotonics as a key ingredient to control the properties of light at the nanoscale, such as the phase,[1,2] near-field amplitude,[3,4] directionality of light scattering,[5,6] spin,[7,8] and orbital angular momentum[9,10] without the necessity of intrinsic material losses as for metal-based approaches. In particular, applications driven by near-field enhancement, such as biomolecular sensing, rely on high resonance quality ($Q = \lambda_{\text{res}}$/FWHM) factors and hence high electromagnetic near-field intensities to achieve maximum specimen sensitivity.[11,12] The inherent correlation between the resonance quality factor and the resonator refractive index[13] known from, e.g., Mie theory, thus led to the advance of all-dielectric nanophotonics based on high refractive index material systems, such as silicon,[14,15] germanium,[16,17] or gallium phosphide.[18,19] Although these materials offer great properties for high-$Q$ resonances in the near-infrared (NIR) and infrared (IR) spectral region, they are accompanied by high material-intrinsic interband absorption losses throughout the visible spectral range due to their intermediate band-gap energies. Owing to these fundamental material limitations, lossless high-index materials throughout the complete visible spectral range are lacking.[20–23] In particular, for the visible wavelength range, there exists a competition between a large band-gap for lossless photonics and a high refractive index of the material for better optical mode confinement and higher $Q$ factors. This physical limitation is mathematically described by the Hervé equation[21]:

$$n = \sqrt{1 + \left(\frac{A}{E_g + B}\right)^2}, \qquad (1)$$

connecting the refractive index $n$ of the material with its band-gap energy $E_g$, where $A$ (= 13.6 eV) and $B$ (= 3.4 eV) are constants. In order to obtain high-$Q$ resonances in the visible (400 - 800 nm) without substantial material losses, a band-gap of 3.54 eV ($\cong$ 350 nm) is required, limiting the maximum refractive index available to $n$ = 2.4 according to Equation 1. In fact, due to the lack of high-index and lossless materials in the visible, experimental demonstrations of high-$Q$ resonances in the blue (i.e., below 460 nm) are still missing.[24] Owing to the low refractive index of the materials, and thus low quality factors of the associated Mie resonances within these materials, this avenue is mostly unexplored.[25,26]

In addition to intrinsic material loss, resonances in nanostructured dielectrics are affected by a radiative loss channel, which becomes dominant when the refractive index contrast between resonator and the environment decreases. Photonic bound states in the continuum (BICs) provide a unique pathway for tailoring radiative losses in addition to the refractive-index-based effects discussed above. Symmetry-broken quasi-BIC (for simplicity BIC) metasurfaces in particular offer precise and straightforward control over the (radiative) $Q$ factor of the resonances via the geometrical asymmetry of the constituent unit cell building blocks,[27,28] which sets them apart from other BIC realizations based on accidental far-field destructive interference[29] or strong mode coupling.[30,31] Initiated by the versatility of the BIC concept, they open up a new avenue for the usage of low-index lossless materials, especially for realizing high-$Q$ resonances throughout the visible spectral range.

Hexagonal boron nitride (hBN),[32] a two-dimensional (2D) van der Waals (vdW) material, exhibits an indirect band-gap of 5.95 eV[33] with an average refractive index of $n = 2.0$, potentially enabling photonic applications down to the ultraviolet (UV) range.[34,35] So far, it was mainly used as insulating or encapsulation layer in electronics and photonics,[36–38] but also found optical applications in a small spectral window (the so-called Reststrahlen band) in the IR due to its strong phononic excitations.[39–42] The few demonstration in the visible have focused on photonic crystal cavities or extended ring resonators,[43,44] which either require suspension or large spatial footprints and have not demonstrated resonances below 580 nm. Nevertheless, its optical properties along with the monocrystalline material quality and compatibility with other 2D materials[45] renders hBN invaluable for the realization of photonic resonances throughout the visible as a basis for enhanced light-matter interaction that has so far been challenging in this spectral range. Beneficially, mechanically exfoliated bulk hBN is obtained in monocrystalline quality, thus diminishing the need for expensive deposition tools and fine engineering of deposition parameters, while simultaneously boosting the quality factors of the resonant modes due to the reduced intrinsic material losses.

In this work, based on a careful analysis of the underlying physics associated with symmetry-broken BICs in low index resonators, we experimentally demonstrate that hBN-based metasurfaces with a low refractive index can support high-$Q$ BIC resonances covering the whole visible spectrum. Our results greatly extend the operational wavelength range for all-dielectric metasurfaces, especially providing high-$Q$ resonances in the blue part of the electromagnetic spectrum for the first time, all within a single sample fabricated from an hBN crystal with constant resonator height. From our numerical investigation, we find that the resonantly enhanced electric near fields are concentrated inside the resonators. Hence, we harness these highly enhanced electromagnetic near fields to demonstrate a 388-fold enhancement for the material-intrinsic SHG in hBN. Because of its combination affinity with other 2D materials, the presence of native single photon sources, and its general optical and chemical properties, our results provide a new paradigm for the development of novel vdW nanophotonic platforms.

**High-Q resonances from low refractive index materials**

Owing to its large band-gap energy and thus low losses in the visible spectral regime, our goal is to obtain sharp resonances from hBN in the complete visible and NIR spectral range (400 to 1000 nm). For the description of the hBN optical properties, we utilize a model based on Sellmaier's equations for both the in-plane $n_i$ and out-of-plane $n_o$ refractive index[46] which accounts for the high material anisotropy. The resulting dispersion for both material axes is depicted on the right-hand side of Figure 1A and is utilized to model the material properties in our numerical calculations.

In order to analyze symmetry-broken BIC metasurfaces composed of hBN, we employ a double rod unit cell. Here, the symmetry breaking is introduced by shortening one rod with respect to the other, such that the geometrical asymmetry is given as $\Delta L = L_2 - L_1$. For the simulation of the transmittance spectra of our metasurfaces, we employ CST microwave studio that allows for the determination of the metasurface response in the frequency domain. In the simulation, we utilize periodic Floquet boundary conditions to calculate the scattering parameters of the metasurface yielding far-field quantities, such as reflectance and transmittance. Furthermore, we assume a plane-wave illumination and set the intrinsic hBN losses to zero within the spectral range of interest. The corresponding numerical spectra are shown on the right-hand side of Figure 1B. The resonance wavelength of the BIC is controlled via fine and continuous scaling of the unit cell with a factor S, which yields closely spaced and sharp resonances throughout the visible to the NIR region.

Enabled by the large band-gap energy, hBN is an ideal candidate to realize high-$Q$ resonances throughout the complete visible spectral range, outperforming other low refractive index materials, such as titanium dioxide or silicon nitride, in the blue part of the visible spectrum. Most importantly, the resonance wavelength of hBN can be scaled to the IR and UV by further adjusting the scaling factor of the unit cell as indicated in numerical simulations in Figure S1 and Figure S2.

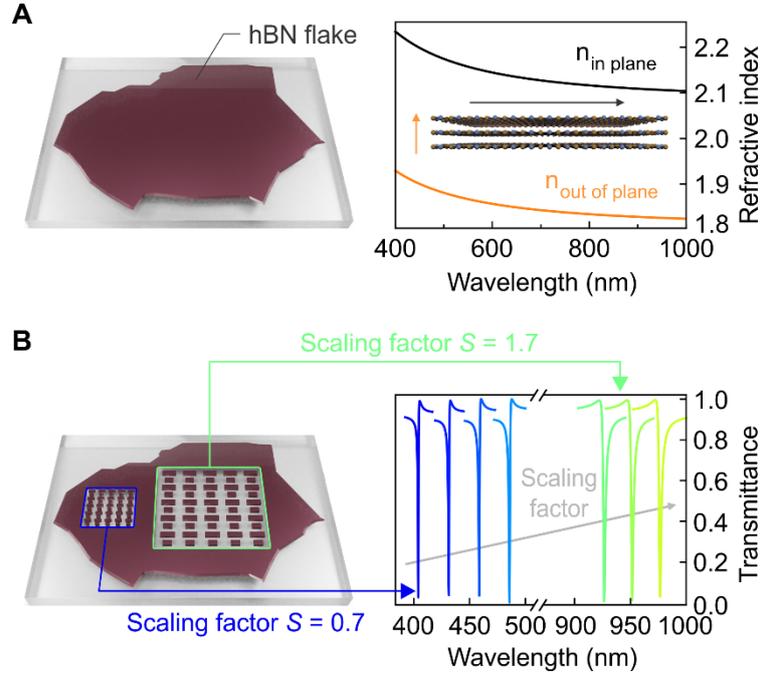

**Figure 1. Hexagonal boron nitride metasurfaces for high-Q resonances in the visible.** (A) Left: schematic illustration of an exfoliated hBN flake on a transparent substrate used for the metasurfaces fabrication. Right: the corresponding refractive index data showing strong anisotropy within the hBN single crystal. (B) Left: schematics of the nanostructured BIC metasurfaces patterned within the hBN. Right: pronounced high-Q resonances from the visible to the NIR by altering the scaling factor of the structures.

In order to demonstrate the feasibility of hBN as photonic resonator material enabled by the BIC concept, we perform in-depth numerical simulations, specifically investigating the influence of the refractive index of the dielectric resonator on the near- and far-field properties of the BIC resonances. We utilize the asymmetric rods BIC unit cell, as introduced above, with fixed geometry settings (periodicities $p_y = p_x + 20$ nm = 430 nm, resonator height $h = 150$ nm, resonator width $w = 100$ nm). We consider resonators without a substrate, to suppress grating modes that could influence the properties of the BIC resonances, and set the resonator material as non-dispersive and lossless. Most importantly, the geometrical asymmetry $\Delta L$ of the system is adjusted to the refractive index $n$ of the resonator material to guarantee the same resonance $Q$ factors for all refractive indices, assuming the generalized asymmetry ($\alpha$) as:

$$\alpha = n \cdot \Delta L. \qquad (2)$$

Uniquely, this $Q$ factor tunability is a defining property of symmetry-broken BIC modes without which the accurate comparison of different resonator materials is infeasible.

A pronounced and spectrally isolated BIC mode can be observed for any resonator refractive index under consideration (Figure 2A). For better comparison, we compare the $n = 4$ index as silicon representative and the $n = 2$ case for hBN in more detail with the corresponding BIC spectra depicted in Figure 2B. These BIC resonances are associated with highly enhanced near-fields (Figure 2C) with the typical BIC pattern of oppositely oscillating dipoles in the *x-y* cross section. The maximum value of the electric near field outside the structures is found to be larger for the Si resonator refractive indices, while the electric field is more distributed and less localized for the case of hBN. By investigating the decay of the normalized electric field $E_{norm}$ away from the structures in z direction, we observe an exponential decay as depicted in Figure 2D for hBN (yellow dots) and silicon (purple dots). By fitting these data with an exponential decay function

$$E_{norm}(z) = \exp\left(-\frac{z}{\zeta}\right), \qquad (3)$$

we extract the decay length $\zeta$ for both resonator materials by determining the distance for which the electric field drops to *1/e* (red dashed line in Figure 2D) with respect to its value at the resonator surface. Indeed, the decay length for hBN-based resonators is roughly twice as large as for silicon-based resonators indicating that the electric near field is more confined to the surface of the resonant structures.

For a deeper understanding of the inherent differences between low index and high index-based BIC metasurfaces, we analyze the decay length as well as the mean and maximum electric fields inside and outside the resonators for a set of different resonator refractive indices as shown in Figure 2E. The investigation of the decay lengths shows a monotonous decrease towards a minimum value around 40 nm for $n = 4$ as obvious from the top panel in Figure 2E, related to the stronger confinements for higher index resonators.

For the investigation of the enhanced electric near-fields, we split the total electric near field in a part inside and outside the resonators and calculate the mean and the maximum value, shown in the middle and bottom panel of Figure 2E. Notably, for increasing refractive indices, the electric fields are pushed outside the resonators as obvious from the higher maximum and mean values of the electric near fields outside the resonators. On the contrary, for smaller refractive indices, the electric field is rather located inside the resonators, rendering them ideally suited for material-intrinsic processes, such as high-harmonic generation or enhanced light-matter interaction utilizing hBN-encapsulated 2D materials. Furthermore, as a consequence of the strong dependence of the electric fields on the $Q$ factor, the BIC approach renders a precise tailoring of the electric fields possible via controlling the geometrical asymmetry of the metasurface. Moreover, BIC metasurfaces based on the low-index materials reduce the requirements on the fabrication accuracy since the same geometrical asymmetries $\Delta L$ lead to lower values of the generalized asymmetry $\alpha$ (see Equation 2).

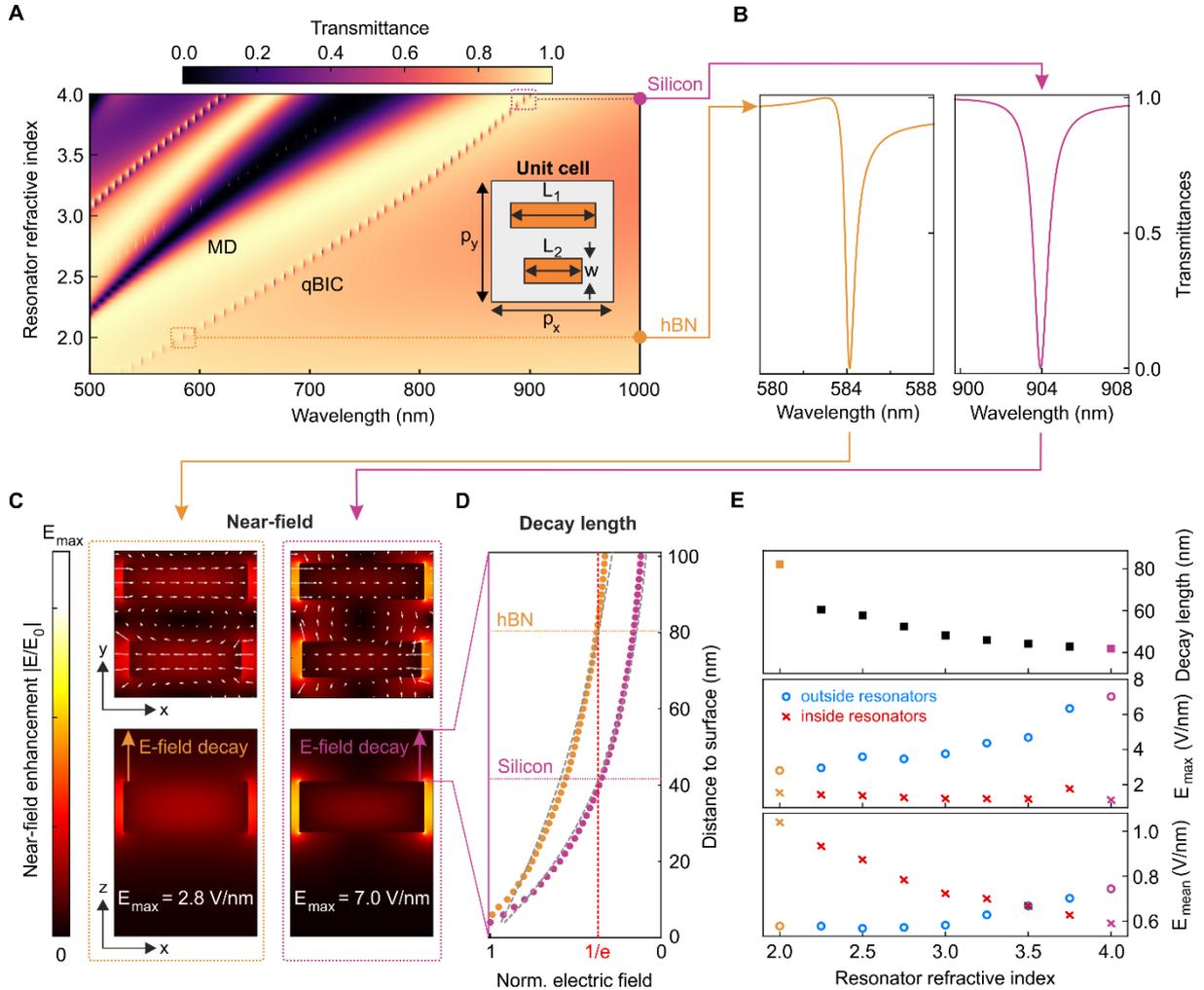

**Figure 2. Impact of the resonator refractive index on the spectral and near-field properties.** (A) Simulated transmittance spectra for the same BIC unit cell with different resonator refractive indices, assuming the refractive index of the substrate $n_{\text{substrate}} = 1$. The resonator materials are assumed to be lossless and without dispersion throughout the spectral range of interest. (B) Representative BIC resonance spectra for refractive indices related to silicon ($n = 4$), in purple, and hBN ($n = 2$), in yellow. (C) Associated BIC near-field enhancement for the silicon and hBN case. (D) Decay of the enhanced electric field away from the surface of the structure as indicated by the arrows in (C). For silicon, the decay length is smaller and the fields are mainly confined to the surface of the resonators when compared to hBN. (E) Upper panel: Decay lengths extracted for the different resonator refractive indices. Maximum value of the electric fields inside the resonators (red crosses) and outside above the substrate (blue circles) in the middle panel and mean value in the lower panel. The behavior of the mean electric fields indicate that the enhanced near-fields are pushed outside the resonators for increasing resonator refractive indices.

It should be noted that the metasurface design has to be carefully optimized, since the spectral separation between the grating mode and the BIC mode decreases. For the experimental situation including a substrate this decrease is most prominent for smaller resonator refractive indices as shown in Figure S3, eventually leading to a vanishing of the BIC mode.

**hBN high-$Q$ resonances through the complete visible range to the NIR**

Owing to this reason, we carefully optimized the design of the unit cell for the experimental demonstration of hBN BIC metasurfaces. We mechanically exfoliate monocrystalline bulk hBN from a bulk single crystal onto a fused silica ($SiO_2$) substrate. Suitable flakes are identified via optical microscopy and their corresponding thickness is determined with a profilometer. We utilize electron beam lithography followed by a gold evaporation step which is used as etching mask. A liftoff and subsequent reactive ion etching transfers the patterns into the hBN flake, before the remaining gold is removed by wet chemistry (see Methods for additional details on the fabrication process). Figure 3A shows the corresponding grey-scale microscope image of a patterned hBN flake, where we fabricated more than 20 arrays, with varying scaling factor, from a single exfoliated crystal.

To characterize our hBN BIC metasurfaces, we illuminate the sample with collimated white light and collect the transmitted signal with a 50x (NA = 0.8) objective. The light is coupled to a multimode fiber and a grating based spectrometer where it is detected by a silicon CCD camera. In order to remove any unwanted spectral features from the sample or the beam path, we reference all our metasurface transmittance spectra to the bare $SiO_2$ substrate.

Figure 3B shows two representative hBN BIC resonances for a scaling factor of $S = 0.65$ (green curve, upper panel) and $S = 1.10$ (brown curve, lower panel) with a spectral separation of almost 300 nm. Uniquely, the BIC resonance is spectrally isolated from any other mode, highly favorable for its usage in sensing applications or light-matter interaction enhancement. As shown in Figure 3D, we demonstrate the full tunability range of the BIC resonances by sweeping the scaling factor $S$ of the unit cell to obtain a comb of well pronounced BIC resonances throughout the complete visible spectrum extending from 413 nm far into the NIR range to 960 nm, only limited by the silicon detector efficiency, depicted in Figure 3E. This resonance scaling corresponds to a tuning range of 1.71 eV, which is, to the best of our knowledge, the broadest BIC-based resonance sweep reported in the literature. Remarkably, the resonant structures are all fabricated from the same flake with a height of 200 nm and the design is optimized for a single wavelength.

To further describe the hBN BIC modes, we utilize a temporal coupled mode theory (TCMT) approach to fit the BIC lineshape and extract the corresponding modulation and $Q$ factors for each scaling factor, as shown in Figure 3C. We find that most BIC resonances show $Q$ factors above 200 with some of them exhibiting modulations exceeding 50 percent in the visible range. Moreover, we observe high-$Q$ resonances below 460 nm, in particular at 413 nm, a spectral position that has not been reported in the literature, with $Q$ factors outperforming all-dielectric- and metal-based approaches by almost one order of magnitude[47,48].

The extracted modulations and $Q$ factors show a maximum range between $S = 0.8$ and $S = 1.1$ while dropping off towards lower or higher scaling factors. We attribute the decreasing $Q$ factors and resonance modulations when approaching the NIR and blue spectral range in the visible to the onset of inter-band absorption in hBN, combined with fabrication intolerances caused by the small geometrical features of the metasurface. Similarly, when pushing the resonances further into the NIR, the optical mode volume is mismatched with the resonator volume and, since the height of the flake is fixed, the grating mode will necessarily spectrally shift towards and over the BIC mode, thus reducing the modulation or leading to a vanishing. Nevertheless, these effects could be minimized by optimizing the hBN flake height and unit cell design for each resonance position.

Beyond the scaling of the optical resonances, we demonstrate the flexibility of the BIC design by controlling the coupling of the BIC to the far-field by varying the geometrical asymmetry of the metasurface unit cell. As shown in Figure 4A, the $Q$ factor increases for smaller geometrical asymmetries until the quasi-BIC transforms into a dark BIC mode without any far-field coupling for $\Delta L = 0$ (black dashed curve), allowing for precise tailoring of the radiative $Q$ factor by controlling the geometrical asymmetry.

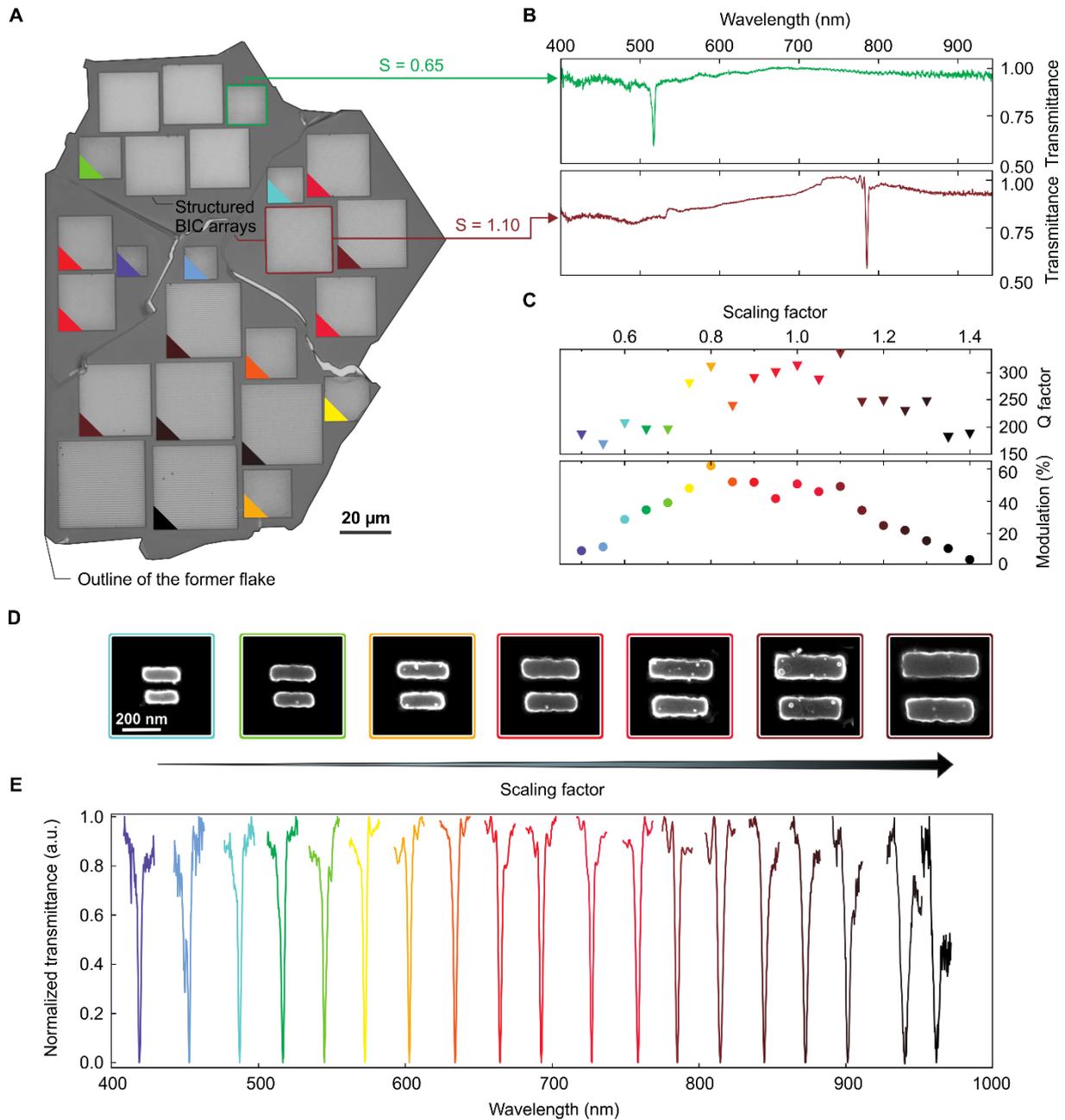

**Figure 3. High-*Q* BIC resonances from hBN through the visible to the NIR.** (A) Optical greyscale micrograph of the bulk hBN flake along with the nanostructured BIC arrays with different scaling factors *S*. (B) Broad range spectra showing the spectral isolation of the BIC mode for *S* = 0.65 (green curve, upper panel) and for *S* = 1.10 (brown curve, lower panel). (C) *Q* factors (top panel) and modulation depths (lower panel) extracted for each resonance scaling factor (shown in panel E) indicate an optimum thickness for the utilized design. (D) SEM micrographs of BIC unit cells with increasing scaling factor. (E) High-*Q* BIC resonances through the visible to the NIR obtained from a single hBN flake by increasing the scaling factor of the metasurface's unit cell.

**SHG enhancement in hBN BIC metasurfaces**

We further combine the versatility of the BIC-based resonances with the highly enhanced electric near fields inside the low-index hBN to demonstrate BIC-enhanced SHG, enabling the spectrally tailored generation of light in the UV spectral region. As shown in Figure 4B, we utilize finite-element method (FEM) simulations (COMSOL Multiphysics) to investigate the SHG efficiency for different geometrical asymmetries. As expected from BIC metasurfaces, higher Q factors are associated with larger near field enhancements leading to a maximum SHG efficiency for the smallest geometrical asymmetry of $\Delta L$ = 10 nm (dark brown curve). Since the experimental $Q$ factors are always limited by fabrication imperfections[49], we choose $\Delta L$ = 75 nm as a trade-off between a high Q factor and a good resonance modulation to guarantee a large SHG enhancement.

The SHG signal in the fabricated sample is excited using a tunable Ti:Sapphire laser and collected in a transmission geometry (see Experimental Section). Figure 4C shows the corresponding SHG signal for an hBN metasurface with $\Delta L$ = 75 nm. We observe a 388-fold BIC-induced SHG enhancement by tuning the excitation source to the two different excitation polarizations, either parallel or perpendicular to the sample. This corresponds to the "on" and "off" states of the BIC resonance, as shown in Figure S4. For incident light which is linearly polarized along the long rod axis ($\mathbf{E}_{par}$, red curve in Figure 4C), the BIC is excited and promotes the SHG enhancement via the highly enhanced near fields. For perpendicularly polarized light ($\mathbf{E}_{per}$, black curve in Figure 4C) the SHG enhancement is absent, since the BIC is not excited in this configuration. This resonance controllability provides an important control mechanism to demonstrate BIC-related enhancement effects and allows us to unambiguously extract an enhancement factor above $10^2$.

As a further confirmation of the BIC-related SHG enhancement, we sweep the excitation wavelength for the same metasurface and observe a clear correlation between the BIC resonance (shown in solid grey in Figure 4D) and the SHG signal at half the BIC resonance for a pump wavelength of 796 nm. By spectrally shifting the excitation wavelength away from the BIC resonance to 830 nm, the SHG signal intensity is largely reduced, underpinning the BIC-resonant enhancement.

We further investigate the dependence of the SHG signal as a function of the incident polarization, shown in Figure 4E for resonant ($\lambda_{pump} = 796\ nm$) and off-resonant ($\lambda_{pump} =$ 830 $nm$) excitation. As expected, on resonance, the SHG signal clearly follows a fully linearly polarized pattern, with the maximum of intensity when the incident light is parallel to the long axis of the rod. On the contrary, when off the BIC resonance, the SHG signal shows no polarization dependence, indicating that the SHG is generated by the intrinsic response of the material, instead of any photonic enhancement.

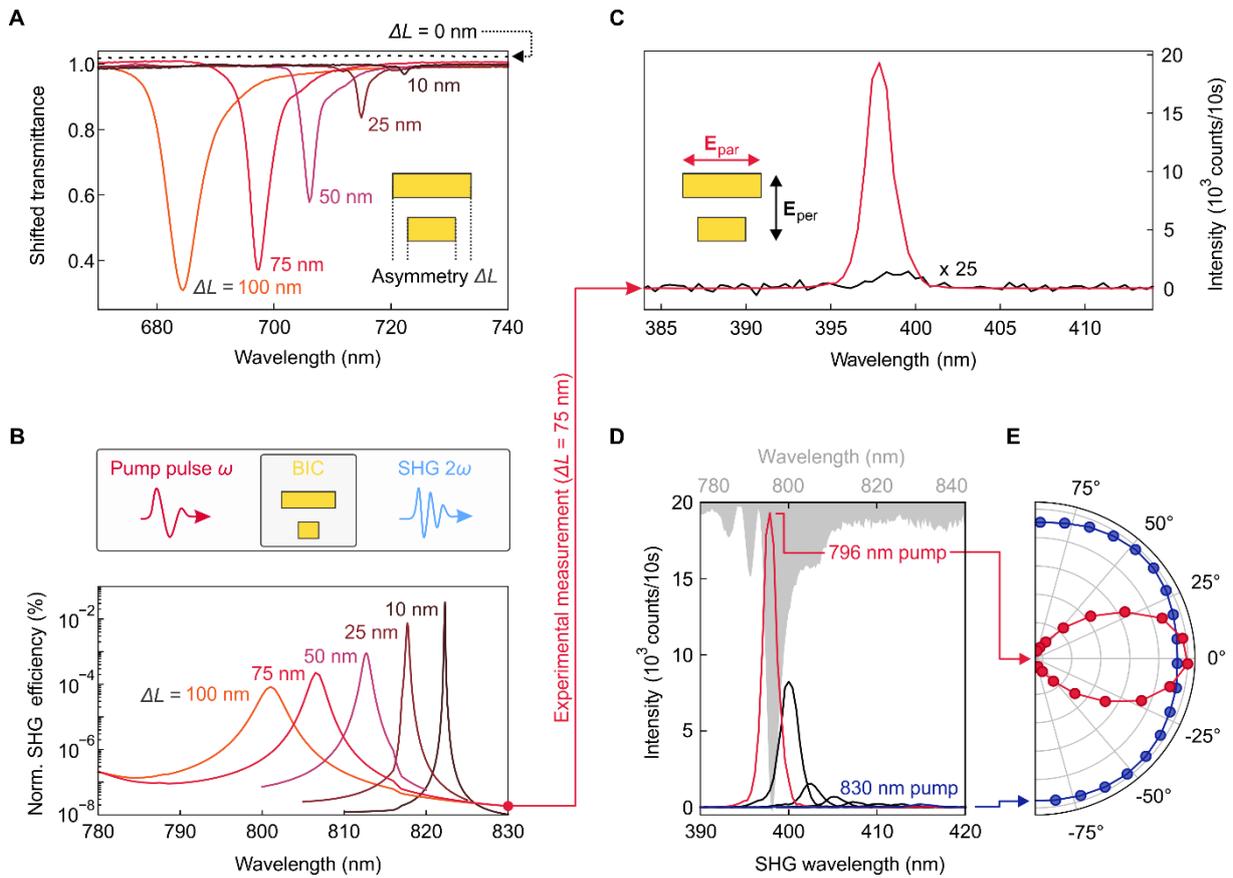

**Figure 4. Symmetry-broken BIC flexibility and SHG enhancement in hBN-based BIC metasurfaces.** (A) Optical white light transmittance spectra for hBN metasurfaces with different asymmetries ranging from $\varDelta L$ = 100 nm (orange curve) to symmetric structures with $\varDelta L$ = 0 nm (black dashed line), showing no coupling to the radiation continuum. (B) Numerical simulations of the normalized SHG efficiency for different geometrical asymmetries, showing the highest enhancement for $\varDelta L$ = 10 nm. (C) SHG signal from a BIC metasurface with $\varDelta L$ = 75 nm, showing a 388-fold BIC-induced enhancement when comparing the two excitation polarizations parallel to the long rod axis ($E_{par}$) and perpendicular to the long rod axis ($E_{per}$). (D) Pump wavelength sweep for the same metasurface as in (C), showing the BIC-resonant enhancement of the SHG. Grey scale: Optical transmittance spectrum of the BIC metasurface with arbitrary units. (E) Normalized SHG intensities for different excitation polarizations showing a full linear polarization signal for the BIC resonance and no detectable polarization for off-resonance pump.

**Conclusion**

In summary, we have numerically investigated and experimentally demonstrated symmetry-broken BIC metasurfaces based on low refractive index monocrystalline hBN. We have analyzed the potential benefits, such as the compatibility of hBN with other materials and the higher electromagnetic fields inside the resonators, but also the drawbacks associated with low index resonators, such as the larger decay length and less field confinement outside the resonators. Furthermore, we outlined the lack of high-index materials for high-$Q$ resonances in the visible rendering low-index materials, such as hBN, invaluable for applications in the visible spectral range. Following our numerical predictions, we realized, for the first time, a broad band resonance sweep of 1.71 eV from the NIR through the complete visible spectral region with well isolated resonances providing $Q$ factors above 150 (with a maximum of 314) and maximum modulations exceeding 50%. Additionally, we showed the lifetime flexibility of the BIC resonances using the geometrical asymmetry of the structures and the selective excitation polarization, giving us full controllability over the resonance position and $Q$ factor from the visible to the IR. Upon resonant excitation, we utilized the highly confined electric near-fields inside the hBN resonators and leveraged the BIC flexibility to demonstrate a 388-fold BIC-based SHG enhancement from hBN in the UV spectral region. Our results not only hold great promise for utilization of hBN as resonator material but also underpin the versatility of BIC metasurfaces for radiative loss management, especially for the technologically relevant visible spectral region. In particular, the hBN-based platform is ideally suited for increased light-matter interaction, such as single photon emission enhancement from bulk hBN or from nanostructured vdW metasurfaces.

**Experimental Section**

**Numerical simulations of hBN-based BIC resonances.** The numerical simulations of the hBN BIC metasurfaces were performed in the frequency domain with CST Microwave using periodic boundary conditions under TE and TM excitation. If considered, the SiO$_2$ substrate was assumed to be lossless and the refractive index was assumed to be isotropic and non-dispersive and was thus set to 1.45. For the modelling of the hBN optical properties, we used Sellmaier's equations according to Ref[46] to account for the strong refractive index anisotropy of hBN. In particular, we used

$$n_\parallel (\lambda) = \sqrt{1 + \frac{3.336\,\lambda^2}{\lambda^2 - 26322}}$$

for the in-plane refractive index and

$$n_\perp (\lambda) = \sqrt{1 + \frac{2.263\,\lambda^2}{\lambda^2 - 26981}}$$

for the out-of-plane refractive index. Furthermore, we assumed the hBN to be lossless ($\kappa = 0$) throughout the spectral region of interest up to 3.4 eV (compared to 6 eV band-gap energy).

**Nanofabrication of hBN BIC metasurfaces**. Bulk hBN exfoliation crystals were obtained via high-pressure growth as described in Ref[50]. Single monocrystalline bulk flakes were isolated from the exfoliation crystal via mechanical exfoliation onto a silicon dioxide (SiO$_2$) substrate with a chromium (Cr) marker system defined before. Suited flakes were identified with an optical microscope and their thicknesses were measured with a Stylus profilometer (Bruker Dektak). Afterwards, a double-layer of the positive-tone electron beam resist poly(methyl methacrylate) (PMMA, Allresist) was spun onto the sample with different chain lengths (80 nm of 950k on top of 100 nm 495k) with soft-baking steps of 3 minutes at 170° C in between. An electrically conductive polymer (Espacer 300Z) was coated on top of the resist to avoid electron charge accumulations and thus pattern distortions. The lithography pattern was defined using electron beam lithography (Raith eLine plus) with an acceleration voltage of 30 kV, aperture size of 15 μm, a working distance of 10 mm, and an area dose of 350 μC/cm$^2$. After the conductive polymer was washed off in a water bath for 10 s, the PMMA double layer was developed in a 3:1 IPA:MIBK (Isopropanol:Methylisobutylketone) solution for 90 s with subsequent 30 s bath in pure IPA. The metal hard-mask consisted of 2 nm titanium (Ti) as adhesion promoter and 60 nm gold (Au) which was deposited via electron beam evaporation. The sample is then left for lift-off overnight at 80 °C in a special remover (Microposit remover 1165). After the lift-off process, the hard mask pattern is transferred into the hBN via inductively coupled reactive ion etching (ICP-RIE) for 50 s using $SF_6$ as precursor gas under 6.0 mTorr pressure with 300 W HF and 150 W ICP power. Finally, the hard mask is removed by placing the sample inside a potassium-based Au etchant and flushing it in water afterwards.

**Optical transmittance measurements of hBN-based BICs.** The optical characterization of the hBN-based BIC metasurfaces was performed using a commercial white light transmission microscopy setup (Witec Alpha series 300). The metasurfaces were excited with linearly polarized collimated white light (Thorlabs OSL2) through the backside of the sample and the collected transmitted intensity was collected with a 50x (NA = 0.8) objective. The collected light was focused into a multi-mode fiber and directed to a grating-based spectrometer, where the light was dispersed and directed onto a silicon CCD. Each spectrum shown in the manuscript was referenced to the bare $SiO_2$ substrate to eliminate unwanted spectral features from the beam path.

**SHG measurements.** The SHG signal was obtained by exciting the sample with the output of a tunable Ti:sapphire laser (Coherent Chameleon Ultra II) with repetition rate of 80 MHz, for the fine tuning of the excitation wavelength to the BIC resonance of the hBN metasurfaces. We employed a transmission geometry where the sample was excited with a low NA objective (10x, NA 0.25) and the SHG signal was collected with a higher NA objective (40x, NA 0.6). The SHG signal was filtered with a shortpass optical filter at 680 nm (Semrock) before being analysed by a grating-spectrometer with a silicon CCD camera (Princeton Instruments). The polarization-resolved SHG signal was obtained by inserting a halfwave plate in the excitation path.

**Supporting Information**

Supporting Information is available from the Wiley Online Library or from the author.

**Conflict of interest**

The authors declare no conflict of interest.

**Acknowledgements**

Our studies were funded by the Deutsche Forschungsgemeinschaft (DFG, German Research Foundation) under grant numbers EXC 2089/1 – 390776260 (Germany´s Excellence Strategy) and TI 1063/1 (Emmy Noether Program), the Bavarian program Solar Energies Go Hybrid (SolTech), the Center for NanoScience (CeNS). K.W. and T.T. acknowledge support from the JSPS KAKENHI (Grant Numbers 19H05790, 20H00354 and 21H05233). L. S. acknowledges funding support through a Humboldt Research Fellowship from the Alexander von Humboldt Foundation. S.A.M. additionally acknowledges the EPSRC ([EP/W017075/1](EP/W017075/1)) and the Lee-Lucas Chair in Physics.

**Data availability**

The data that support the findings of this study are available from the corresponding author upon reasonable request.

**References**

1. Lawrence, M. *et al.* High quality factor phase gradient metasurfaces. *Nat. Nanotechnol.* **15**, 956–961 (2020).
2. Overvig, A. C. *et al.* Dielectric metasurfaces for complete and independent control of the optical amplitude and phase. *Light Sci. Appl.* **8**, 92 (2019).
3. Tittl, A. *et al.* Imaging-based molecular barcoding with pixelated dielectric metasurfaces. *Science* **360**, 1105–1109 (2018).
4. Yesilkoy, F. *et al.* Ultrasensitive hyperspectral imaging and biodetection enabled by dielectric metasurfaces. *Nat. Photonics* **13**, 390–396 (2019).
5. Shibanuma, T. *et al.* Experimental Demonstration of Tunable Directional Scattering of Visible Light from All-Dielectric Asymmetric Dimers. *ACS Photonics* **4**, 489–494 (2017).
6. Karst, J. *et al.* Electrically switchable metallic polymer nanoantennas. *Science* **374**, 612–616 (2021).
7. Bao, Y. *et al.* On-demand spin-state manipulation of single-photon emission from quantum dot integrated with metasurface. *Sci. Adv.* **6**, 1–8 (2020).
8. Huo, P. *et al.* Photonic Spin-Multiplexing Metasurface for Switchable Spiral Phase Contrast Imaging. *Nano Lett.* **20**, 2791–2798 (2020).
9. Ren, H. *et al.* Metasurface orbital angular momentum holography. *Nat. Commun.* **10**, 2986 (2019).

10. Ren, H. *et al.* Complex-amplitude metasurface-based orbital angular momentum holography in momentum space. *Nat. Nanotechnol.* **15**, 948–955 (2020).

11. Wang, J. *et al.* All-Dielectric Crescent Metasurface Sensor Driven by Bound States in the Continuum. *Adv. Funct. Mater.* **31**, 2104652 (2021).

12. Yang, Y., Kravchenko, I. I., Briggs, D. P. & Valentine, J. All-dielectric metasurface analogue of electromagnetically induced transparency. *Nat. Commun.* **5**, 5753 (2014).

13. Baranov, D. G. *et al.* All-dielectric nanophotonics: the quest for better materials and fabrication techniques. *Optica* **4**, 814 (2017).

14. Staude, I. & Schilling, J. Metamaterial-inspired silicon nanophotonics. *Nat. Photonics* **11**, 274–284 (2017).

15. Wu, C. *et al.* Spectrally selective chiral silicon metasurfaces based on infrared Fano resonances. *Nat. Commun.* **5**, 3892 (2014).

16. Leitis, A. *et al.* Angle-multiplexed all-dielectric metasurfaces for broadband molecular fingerprint retrieval. *Sci. Adv.* **5**, eaaw2871 (2019).

17. Campione, S. *et al.* Broken Symmetry Dielectric Resonators for High Quality Factor Fano Metasurfaces. *ACS Photonics* **3**, 2362–2367 (2016).

18. Anthur, A. P. *et al.* Continuous Wave Second Harmonic Generation Enabled by Quasi-Bound-States in the Continuum on Gallium Phosphide Metasurfaces. *Nano Lett.* **20**, 8745–8751 (2020).

19. Hüttenhofer, L. *et al.* Metasurface Photoelectrodes for Enhanced Solar Fuel Generation. *Adv. Energy Mater.* **11**, 2102877 (2021).

20. Moss, T. S. Relations between the Refractive Index and Energy Gap of Semiconductors. *Phys. status solidi* **131**, 415–427 (1985).

21. Hervé, P. & Vandamme, L. K. J. General relation between refractive index and energy gap in semiconductors. *Infrared Phys. & Technol.* **35**, 609–615 (1994).

22. Khurgin, J. B. Expanding the Photonic Palette: Exploring High Index Materials. *ACS Photonics* **9**, 743–751 (2022).

23. Shim, H., Monticone, F. & Miller, O. D. Fundamental Limits to the Refractive Index of Transparent Optical Materials. *Adv. Mater.* **33**, 2103946 (2021).

24. Sun, S. *et al.* All-Dielectric Full-Color Printing with TiO 2 Metasurfaces. *ACS Nano* **11**, 4445–4452 (2017).

25. Zhang, C. *et al.* Low-loss metasurface optics down to the deep ultraviolet region. *Light Sci. Appl.* **9**, 55 (2020).

26. Kim, K.-H. & Kim, I.-P. Quasi-bound states in the continuum with high Q -factors in

metasurfaces of lower-index dielectrics supported by metallic substrates. *RSC Adv.* **12**, 1961–1967 (2022).

27. Koshelev, K., Lepeshov, S., Liu, M., Bogdanov, A. & Kivshar, Y. Asymmetric Metasurfaces with High-Q Resonances Governed by Bound States in the Continuum. *Phys. Rev. Lett.* **121**, 193903 (2018).

28. Azzam, S. I. & Kildishev, A. V. Photonic Bound States in the Continuum: From Basics to Applications. *Adv. Opt. Mater.* **9**, 2001469 (2021).

29. Kodigala, A. *et al.* Lasing action from photonic bound states in continuum. *Nature* **541**, 196–199 (2017).

30. Rybin, M. V. *et al.* High-Q Supercavity Modes in Subwavelength Dielectric Resonators. *Phys. Rev. Lett.* **119**, 243901 (2017).

31. Koshelev, K. *et al.* Subwavelength dielectric resonators for nonlinear nanophotonics. *Science* **367**, 288–292 (2020).

32. Gil, B., Cassabois, G., Cusco, R., Fugallo, G. & Artus, L. Boron nitride for excitonics, nano photonics, and quantum technologies. *Nanophotonics* **9**, 3483–3504 (2020).

33. Cassabois, G., Valvin, P. & Gil, B. Hexagonal boron nitride is an indirect bandgap semiconductor. *Nat. Photonics* **10**, 262–266 (2016).

34. Ferreira, F., Chaves, A. J., Peres, N. M. R. & Ribeiro, R. M. Excitons in hexagonal boron nitride single-layer: a new platform for polaritonics in the ultraviolet. *J. Opt. Soc. Am. B* **36**, 674 (2019).

35. Xu, H. *et al.* Magnetically tunable and stable deep-ultraviolet birefringent optics using two-dimensional hexagonal boron nitride. *Nat. Nanotechnol.* 1–1 (2022) doi:10.1038/s41565-022-01186-1.

36. Dean, C. R. *et al.* Boron nitride substrates for high-quality graphene electronics. *Nat. Nanotechnol.* **5**, 722–726 (2010).

37. Haigh, S. J. *et al.* Cross-sectional imaging of individual layers and buried interfaces of graphene-based heterostructures and superlattices. *Nat. Mater.* **11**, 764–767 (2012).

38. Liu, Y. *et al.* Van der Waals heterostructures and devices. *Nat. Rev. Mater.* **1**, 16042 (2016).

39. Autore, M. *et al.* Boron nitride nanoresonators for phonon-enhanced molecular vibrational spectroscopy at the strong coupling limit. *Light Sci. Appl.* **7**, 17172–17172 (2018).

40. Autore, M. *et al.* Enhanced Light–Matter Interaction in 10 B Monoisotopic Boron Nitride Infrared Nanoresonators. *Adv. Opt. Mater.* **9**, 2001958 (2021).

41. Caldwell, J. D. *et al.* Photonics with hexagonal boron nitride. *Nat. Rev. Mater.* **4**, 552–567 (2019).

42. Caldwell, J. D. *et al.* Sub-diffractional volume-confined polaritons in the natural hyperbolic material hexagonal boron nitride. *Nat. Commun.* **5**, 5221 (2014).

43. Kim, S. *et al.* Photonic crystal cavities from hexagonal boron nitride. *Nat. Commun.* **9**, 2623 (2018).

44. Fröch, J. E., Hwang, Y., Kim, S., Aharonovich, I. & Toth, M. Photonic Nanostructures from Hexagonal Boron Nitride. *Adv. Opt. Mater.* **7**, 1–6 (2019).

45. Geim, A. K. & Grigorieva, I. V. Van der Waals heterostructures. *Nature* **499**, 419–425 (2013).

46. Rah, Y., Jin, Y., Kim, S. & Yu, K. Optical analysis of the refractive index and birefringence of hexagonal boron nitride from the visible to near-infrared. *Opt. Lett.* **44**, 3797 (2019).

47. Zhu, X. *et al.* Beyond Noble Metals: High Q -Factor Aluminum Nanoplasmonics. *ACS Photonics* **7**, 416–424 (2020).

48. Semmlinger, M. *et al.* Vacuum Ultraviolet Light-Generating Metasurface. *Nano Lett.* **18**, 5738–5743 (2018).

49. Kühne, J. *et al.* Fabrication robustness in BIC metasurfaces. *Nanophotonics* **10**, 4305–4312 (2021).

50. Taniguchi, T. & Watanabe, K. Synthesis of high-purity boron nitride single crystals under high pressure by using Ba–BN solvent. *J. Cryst. Growth* **303**, 525–529 (2007).

# Supporting information for
# High-Q nanophotonics over the full visible spectrum enabled by hexagonal boron nitride metasurfaces

Lucca Kühner[1], Luca Sortino[1], Benjamin Tilmann[1], Thomas Weber[1], Kenji Watanabe[2], Takashi Taniguchi[3], Stefan A. Maier[4,5,1], and Andreas Tittl[1,*]

[1]Chair in Hybrid Nanosystems, Nanoinstitute Munich, and Center for NanoScience, Faculty of Physics, Ludwig-Maximilians-University Munich, Königinstraße 10, 80539 München, Germany

[2]Research Center for Functional Materials, National Institute for Materials Science, 1-1 Namiki, Tsukuba 305-0044, Japan

[3]International Center for Materials Nanoarchitectonics, National Institute for Materials Science, 1-1 Namiki, Tsukuba 305-0044, Japan

[4]School of Physics and Astronomy, Monash University, Wellington Rd, Clayton VIC 3800, Australia

[5]The Blackett Laboratory, Department of Physics, Imperial College London, London, SW7 2AZ, United Kingdom

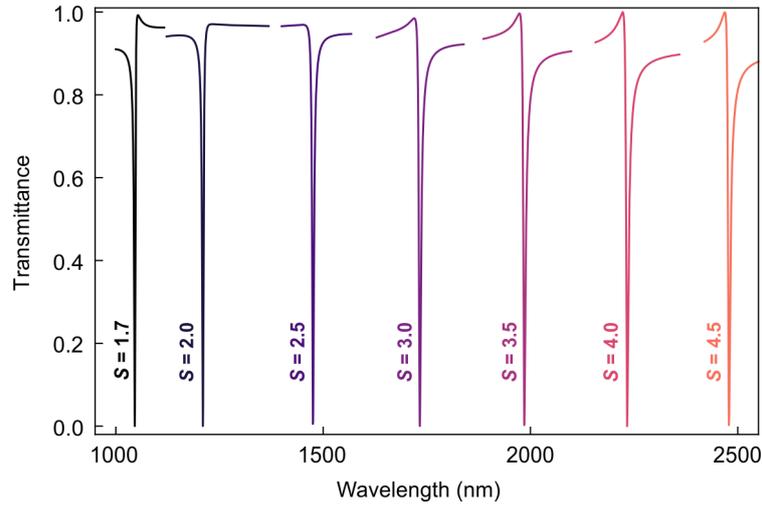

**Figure S1. Scaling factor (*S*) sweep of the metasurface into the IR.** Simulations for a thicker flake with $h$ = 400 nm, showing pronounced BIC resonances from 1000 nm to 2500 nm. The unit cell dimensions are the same as for Figure 2 except a slightly reduced periodicity of $p_x$ = 390 nm. The simulations are performed on a $SiO_2$ substrate assuming a dispersion less refractive index of $n$ = 1.45.

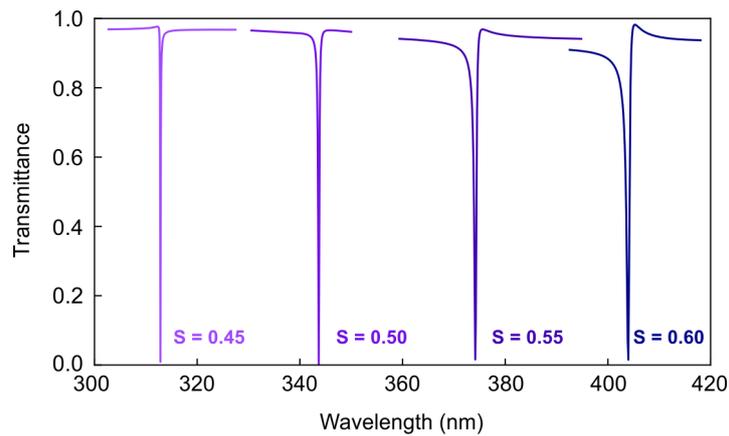

**Figure S2. Scaling factor (*S*) sweep of the metasurface in the UV region.** The unit cell dimensions are the same as for Figure 2 except a slightly reduced periodicity of $p_x$ = 390 nm with a height of $h$ = 200 nm. The simulations are performed on a $SiO_2$ substrate assuming a dispersion less refractive index of $n$ = 1.45.

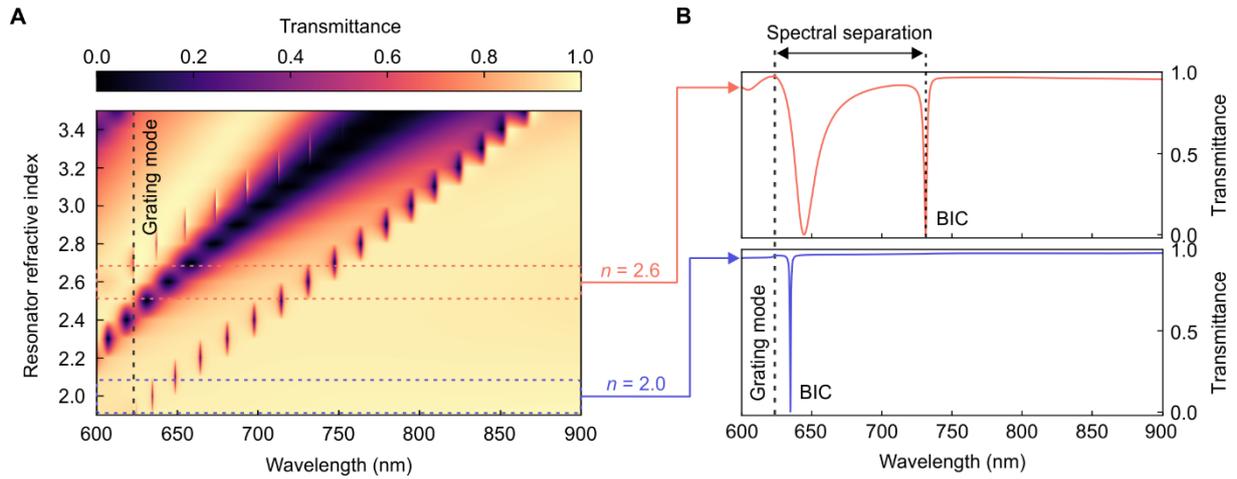

**Figure S3. Influence of the resonator refractive index on the spectral separation of the grating mode and the BIC.** (A) Transmittance spectra for different resonator refractive indices showing the shifting of the BIC towards the grating mode for smaller refractive indices. (B) Two representative simulated transmittance spectra for $n = 2.0$ (blue) and $n = 2.6$ (orange) make the different spectral separation between the BIC and the grating mode clearer.

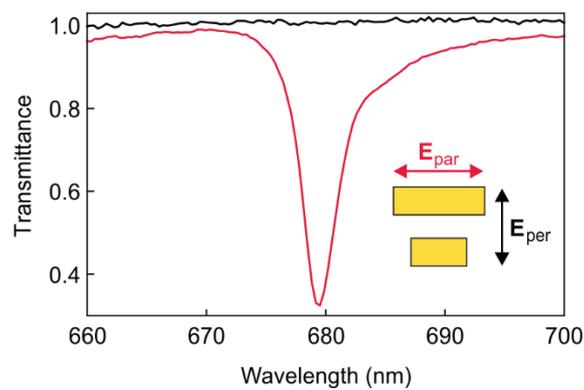

**Figure S4. Resonance controllability via the excitation polarization.** The BIC mode can be controlled via simply choosing the linear excitation polarization parallel to the long rod axis ($\mathbf{E}_{par}$) or perpendicular to it ($\mathbf{E}_{per}$).

23